\documentclass[12pt]{article}%
\pdfoutput=1 

\usepackage{natbib}
\usepackage{amsmath}
\usepackage{amsbsy}
\usepackage{amscd}
\usepackage{amsopn}
\usepackage{amstext}
\usepackage{amsxtra}
\usepackage{amssymb}
\usepackage{amsthm,amsfonts}
\usepackage{graphicx}
\usepackage{epstopdf}

\usepackage{epsfig}

\def\lesssim{\mathrel{\hbox{\rlap{\hbox{\lower4pt\hbox{$\sim$}}}\hbox{$<$}}}}

\def\Cv{\mathcal{V}}
\def\Cl{\mathcal{L}}
\def\Cm{\mathcal{M}}
\def\Cw{\mathcal{W}}

\def\Cs{\mathcal{S}}
\def\be{\begin{equation}}

\def\indeg{\mbox{deg}^-}
\def\outdeg{\mbox{deg}^+}

\def\beq{\begin{eqnarray}}
\def\eeq{\end{eqnarray}}
\def\beqq{\begin{eqnarray*}}
\def\eeqq{\end{eqnarray*}}
\def\beeq{\begin{eqnarray*}}
\def\eeeq{\end{eqnarray*}}
\def\be{\begin{equation}}
\def\ee{\end{equation}}

\newtheorem{theorem}{Theorem}
\newtheorem{lemma}[theorem]{Lemma}
\newtheorem{proposition}[theorem]{Proposition}

\newtheorem{remark}[theorem]{Remark}

\newtheorem{definition}[theorem]{Definition}

\makeindex             

\begin{document}

\title{A framework for analyzing contagion in banking networks}
\author{T. R.  Hurd$^1$,  James P. Gleeson$^2$\\
\emph{ $^1$ Department of Mathematics, McMaster University, Canada }\\
 \emph{$^2$MACSI, Department of Mathematics \& Statistics,}\\\emph{ University of Limerick, Ireland}}
\date{October 17, 2011}%
%
\maketitle


\abstract{A probabilistic framework is introduced that represents stylized banking networks and aims to predict the size of contagion events. In contrast to previous work on random financial networks, which assumes independent connections between banks, the possibility of disassortative edge probabilities (an above average tendency for small banks to link to large banks) is explicitly incorporated.
We give a probabilistic analysis of the default cascade triggered by shocking the network. We find that the cascade can be understood as an explicit iterated mapping on a set of edge probabilities that converges to a fixed point.  A cascade condition is derived that characterizes whether or not an infinitesimal shock to the network can grow to a finite size cascade, in analogy to the basic reproduction number $R_0$ in epidemic modeling. It provides an easily computed measure of the systemic risk inherent in a given banking network topology. An analytic formula is given for the frequency of global cascades, derived from percolation theory on the random network. Two simple examples are used to demonstrate that edge-assortativity can have a strong effect on the level of systemic risk as measured by the cascade condition.  Although the analytical methods are derived for infinite networks, large-scale Monte Carlo simulations are presented that demonstrate the applicability of the results to finite-sized networks. Finally, we propose a simple graph theoretic quantity, which we call ``graph-assortativity'', that seems to best capture systemic risk.}
\bigskip


\newpage

\section{Introduction}
\label{sec:1}
The study of contagion in financial systems is very topical in light of the recent global credit crisis and the resultant damage inflicted on financial institutions. ``Contagion'' refers to the spread of defaults through a  system of financial institutions, with each successive default causing increasing pressure on the remaining components of the system. The term ``systemic risk'' refers to the contagion-induced threat to the financial system as a whole, due to the default of one (or more) of its component institutions.

It is widely held that financial systems (see \cite{Upper11} and references therein), defined for example as the collection of banks and financial institutions in a developed country, can be modelled as a random network of {\it nodes} or {\it vertices} with stylized balance sheets,  connected by directed links or edges that represent exposures (``interbank loans''), each edge with a positive weight that represents the size of the exposure. If ever a node becomes ``insolvent'' and ceases to operate as a bank, it will create balance sheet shocks to other nodes, creating the potential of chains of insolvency that we will call ``default cascades''.
Financial networks are difficult to observe because interbank data is often not publicly available, but studies have indicated that they  share characteristics of other types of technological and social networks, such as the world wide web and Facebook. For example, the degree distributions of financial networks are thought to be ``fat-tailed'' since a significant number of banks are very highly connected. A less studied feature observed in financial networks (and as it happens, also the world wide web) is that they are highly ``disassortative'' (see \cite{Newman02}). This refers to the property that any bank's counterparties (i.e. their graph neighbours) have a tendency to be banks of an opposite character. For example, it is observed that small banks tend to link preferentially to large banks rather than other small banks. Commonly, social networks are observed to be assortative rather than disassortative. Structural characteristics such as degree distribution and assortativity are felt to be highly relevant to the propagation of contagion in networks but the nature of such relationships is far from clear.

Our aim here is to develop a mathematical framework that will be able to determine the systemic susceptibility in a rich class of infinite random network  models with enough flexibility to include the most important structural characteristics of real financial networks, in particular with general degree distributions and a prescribed degree of edge-assortativity. Our starting point will be the Gai and Kapadia (hereafter referred to as GK for short) cascade model of \cite{GaiKapa10} and the analytical methods developed there and in \cite{GleHurMelHac11} for this model.  The basic assumptions introduced in the GK model are:
\begin{enumerate}
  \item The network is a large (actually infinite) random directed graph with a prescribed degree distribution;
  \item Each node is labelled with a stylized banking balance sheet that identifies its external assets and liabilities, its internal (i.e.  total interbank) assets and liabilities, and $\gamma$, its net worth or equity (i.e. its total assets minus its total liabilities). Initially, the system is in equilibrium, meaning each node has positive net worth $\gamma>0$.
  \item Each directed edge is labelled with a deterministic weight that represents the positive exposure of one bank to another. These weights depend deterministically on the in-degree of the edge, and are consistent with the interbank (IB) assets and liabilities at each node;
  \item A random shock is applied to the balance sheets in the system that triggers the default or insolvency of a fixed fraction of nodes;
  \item The residual value available to creditors of a defaulted bank is zero, and thus the shock has the potential to trigger a cascade of further bank defaults.
\end{enumerate}
The principle of {\it limited liability} for banks means that equity holders are never asked to cover a negative net worth of an insolvent firm. Instead, the insolvent firm is assumed to ``default'', meaning it ceases to operate as a going concern, and its creditors divide the residual value. Since this residual value is always less than the nominal liabilities, creditor banks thus receive a shock to their balance sheets, creating the potential for a default cascade.  The GK model makes a very simple {\it zero recovery} assumption that residual values of defaulted banks will be zero, and thus every time a bank defaults a maximal possible shock will be transmitted to its creditors.

The framework for random financial networks presented here generalizes the GK model in one important respect, namely that the edge degree distribution is arbitrary, allowing for any desired amount of assortativity or disassortativity in the network. Under these more general assumptions, we are able to prove our main result about default cascades, which draws the following conclusions:
\begin{enumerate}
  \item The probability that an edge is defaulted after $n$ steps of the cascade depends on the in-degree $j$ of the edge, but not on the out-degree. Let these quantities be denoted $\vec{a}^{(n)}=\{a^{(n)}_j\}_{j\in \mathbb{Z}_+}, \mathbb{Z}_+:=\{0, 1,2,\dots\}$ for each value of the in-degree $j$.
  \item The sequence $\{\vec{a}^{(n)}\}_{n=0,1,\dots}$ satisfies a recursion
    \[\vec{a}^{(n+1)}=G(\vec{a}^{(n)}), \quad n=0,1,\dots\]
where the monotonically increasing mapping $G:[0,1]^{\mathbb{Z}_+}\to [0,1]^{\mathbb{Z}_+}$  depends explicitly on the structure of the network and the initial shock distribution.
\end{enumerate}
An earlier result of this type was proved in \cite{GleHurMelHac11} in a situation in which the probabilities $a$ are independent of $j$, and $G$ reduces to a scalar function.

One important implication of the above result is the so-called ``cascade condition'' which is a statement about the derivative of $G$ computed in the limit as the initial shock goes to zero. It represents the condition that a tiny initial shock can create a finite size cascade, and is the main analytical quantity that determines the stability of the financial network. For example, with a given network structure, we are able to compute a critical value $\gamma_c$  of the net worth parameter that separates a no-cascade region from a cascade region.

The remainder of this paper is structured as follows. In Section~\ref{sec:GK} the extended GK model is described in detail, and the analytical description of its solution is given in  Section~\ref{sec:cascade}. Section~\ref{sec:casccond} discusses the cascade condition, and Section~\ref{Frequency} derives a formula for the frequency of large scale cascades arising from an infinitesimally small seed. Numerical Monte Carlo simulation results and the corresponding analytical predictions  are compared for a simple class of networks in Section~\ref{sec:num}. This section also investigates in more detail the relation between assortativity and systemic risk in a richer family of infinite network examples.  Section~\ref{sec:conclusion} concludes.

 \section{The Extended GK Model}\label{sec:GK}
 In this section we completely specify the extended GK modelling framework for which our analytical techniques will apply. The specification will consist of three levels: first, the random directed graph model for the ``skeleton'' of the network; second, a specification of balance sheet values corresponding to all nodes and edges; thirdly, a specification of the type of initial shocks that will be considered. We shall work within a probability space $(\Omega_N,\mathcal{F}_N,\mathbb{P})$ for some $N\le \infty$, where a general outcome $\omega\in\Omega_N$ is a directed graph with $N$ nodes, with specified balance sheet values and an initial shock.

 \subsection{The Skeleton Network}
\label{sec:skel}
 The first step in building a random financial network is to build the skeleton directed graph that labels the banks and their interbank connections.  Our construction is an extension of the Erd\"os-Renyi random graph model, and to describe it we introduce certain graph theoretic definitions and notation:
\begin{enumerate}
  \item A node $v\in\Cv=\cup_{jk}\Cv_{jk}$ has type $(j,k)$ means its in-degree is $\indeg(v)=j$ and its out-degree is $\outdeg(v)=k$. We shall write $k_v=k,j_v=j$ for any $v\in\Cv_{jk}$ and allow degrees to be any non-negative integer.
  \item An edge $\ell\in\Cl=\cup_{kj}\Cl_{kj}$ is said to have type $(k,j)$ with in-degree $j$ and out-degree $k$ if it is an out-edge of a node with out-degree $k$ and an in-edge of a node with in-degree $j$. We shall write $\outdeg(\ell)=k_\ell=k$ and $\indeg(\ell)=j_\ell=j$ for any $ \ell\in\Cl_{kj}$.
  \item We write $\Cl^+_v$  (or $\Cl^-_v$) for the set of out-edges (respectively, in-edges) of a given node $v$ and $v^+_\ell$ (or $v^-_\ell$) for the node for which $\ell$ is an out-edge (respectively, in-edge).
  \item We will always write  $j,j',j''$, etc. to refer to in-degrees while $k,k',k''$, etc. refer to out-degrees.
\end{enumerate}

\begin{figure}[h]
\centering
\includegraphics[scale=.7]
{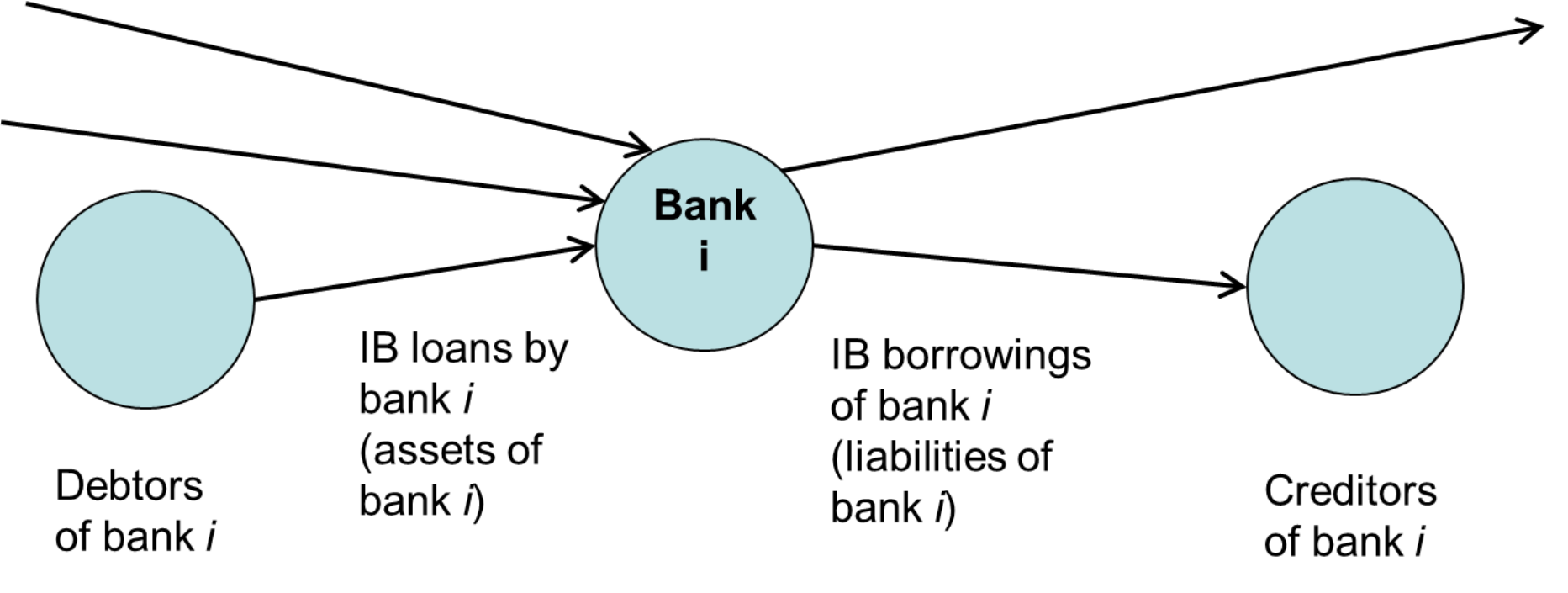}
\caption{The skeleton structure of the network locality of a bank $v$. Bank $v$ is in the $(j,k)=(3,2)$ class, since it has 3 debtors and 2 creditors in the interbank (IB) network.} \label{figS1}
\end{figure}

Figure \ref{figS1} illustrates the neighbourhood of a type $(3,2)$ bank. Our random graph model is now characterized by a certain independence structure, and  by probability laws $P, Q$ for nodes and edges.
\begin{definition}\begin{enumerate}
  \item For each $j,k$,
$P_{jk}:=\mathbb{P}[\Cv_{jk}]$ is the probability of a type $(j,k)$ node. This distribution has marginals  $P^{+}_k:=\sum_j P_{jk}, P^{-}_j:=\sum_kP_{jk}$.
  \item For each $j,k$,
$Q_{kj}:=\mathbb{P}[\Cl_{kj}]$ is the probability of a type $(k,j)$ edge.
  This distribution has marginals  $Q^{+}_k:=\sum_j Q_{kj}, Q^{-}_j:=\sum_kQ_{kj}$.
  \end{enumerate}
  \end{definition}

For any finite number of nodes $N$, a random graph with the required statistics can be constructed by following a generalization of the ``configuration graph'' construction used to construct Erd\"os-Renyi graphs:
  \begin{enumerate}
  \item First, $N$ degree pairs $(j_n,k_n), n=0, 1,\dots, N$ are drawn independently from the $P$ distribution. The $n$th node then has $j_n$ ``in-stubs'' and $k_n$ ``out-stubs''.
    \item Next, for each in-stub attached to a node with degree pair $(j,k), j>0$ one draws an out-degree $k'>0$ from the conditional distribution\footnote{Recall the definition of conditional probability for any pair of sets of outcomes $A,B$:
    \[ \mathbb{P}[A|B]=\mathbb{P}[A\cap B]/\mathbb{P}[B]\ .\]}
    $\mathbb{P}[k_\ell=k|j_\ell=j]=Q_{kj}/Q^{-}_j$: we can say the type of this in-stub is $(k',j)$. The collection of such draws is mutually independent.
  \item Then, each in-stub of type $(k,j)$ in succession is glued to a random out-stub selected with uniform probability from all remaining unpaired out-stubs with out-degree $k$. Again, the collection of such draws is mutually independent.
  \item Finally, the graph is accepted if all in-stubs and out-stubs are paired, and rejected if not. 
\end{enumerate}

For a realization of the above finite $N$ graph construction to be accepted, it is necessary for the total number of in-stubs with any given out-degree $k$ to equal the number of out-stubs connected to nodes with this value of $k$. As $N\to\infty$, this leads by the law of large numbers to a consistency condition relating the marginals of $Q$ to the marginals of $P$.  Some further thought about the wiring of in-stubs to out-stubs leads one to deduce certain independence properties of the random graph in the $N=\infty$ limit. Heuristically, any node $v$ or edge $\ell$ divides the graph into   ``downstream'' and ``upstream'' parts. We can say that any downstream probability is independent of the upstream graph and vice versa. Based on these considerations, we adopt the following description of the random directed graph model:
 \begin{definition}
For any $N>1$ (including $N=\infty$), let  $P_{jk}, Q_{kj}, j,k=0,1,\dots$ be  collections of node and edge probabilities, subject to the constraints
\[Q^{+}_k=kP^{+}_k/z, \quad Q^{-}_j=jP^{-}_j/z\]
for all $j,k$, where the mean degree is $z=\sum_k kP^{+}_k=\sum_j j P^{-}_j$. For $N<\infty$, the random directed graph model ${\cal G}(N,P,Q)$ is defined by the above configuration graph construction. For $N=\infty$,
 the random directed graph model ${\cal G}(N=\infty, P,Q)$ is instead characterized by the assumption of independence of downstream and upstream probabilities, conditioned on the type of node or edge. This independence condition is illustrated by the important case\footnote{A conditional independence structure more general than \eqref{decomposition2}, not arising from the above graph construction, is analyzed in  \cite{BoguAnge05}.}: For all $j,j',j'',k,k',k''$,
\be
\mathbb{P}[v^+_\ell\in\Cv_{jk},v^-_\ell\in\Cv_{j'k'}|\ell\in\Cl_{k''j''}]=\delta_{j'j''}\delta_{kk''}\mathbb{P}[j_v=j|k_v=k]\ \mathbb{P}[k_v=k'|j_v=j']\ .\label{decomposition2}
\ee
\end{definition}

%
%
%
%
%
%

  \begin{remark} {\bf (Independent edge condition)} The special case $Q_{kj}=kjP^{-}_jP^{+}_k/z^2=Q^{-}_jQ^{+}_k$ arises when each in-stub is glued to an out-stub selected uniformly from the collection of all un-paired out-stubs, and is the usual notion of random graph. We are interested in the more general case described above because observed financial networks do not appear to satisfy the independent edge condition. Apparently real financial networks have the ``edge-disassortative property'' that high degree banks attach preferentially to low degree banks.
\end{remark}

  \begin{remark} A natural measure of edge-assortativity by degree is the ``edge-assortativity coefficient'' $r_Q\in[-1,1]$ given by
  \be\label{edgeassortcoef}
  r_Q=\frac{\sum_{jk}jk[Q_{kj}-Q^-_jQ^+_k]}{\sqrt{\left(\sum_{j}j^2Q^-_j-(\sum_{j}jQ^-_j)^2\right)\left(\sum_{k}k^2Q^+_k-(\sum_{k}kQ^+_k)^2\right)}}\ .
  \ee
 This quantity has been measured for the Brazilian banking network in \cite{ContMousSant10} and found to be usually negative.  However, we will find some evidence that systemic risk of a network is related more strongly to a combination of edge-assortativity and  node-assortativity (arising from the dependence between in- and out- degrees of nodes). We therefore define a measure we call the ``graph-assortativity coefficient'' $r\in[-1,1]$ given by
  \be\label{graphassortcoef}
  r=\frac{\sum_{jj'}jj'[B_{jj'}-B^-_jB^+_{j'}]}{\sqrt{\left(\sum_{j}j^2B^-_j-(\sum_{j}jB^-_j)^2\right)\left(\sum_{j'}{j'}^2B^+_{j'}-(\sum_{j'}j'B^+_{j'})^2\right)}}\ .
  \ee
  where $B_{jj'}$ is the joint distribution of the in-degree of pairs of nodes connected by an edge:
  \beqq B_{jj'}&=&\mathbb{P}[j_v=j,j_{v'}=j'|\mbox{$v$ is joined by a single out-edge $\ell$ to $v'$}]\\
  &=&\sum_k\frac{P_{jk}Q_{kj'}}{P^+_k}\ ,
  \eeqq
and $B^-_j=\sum_{j'}B_{jj'}, B^+_{j'}=\sum_j B_{jj'}$ are the marginals. Here, the formula for $B_{jj'}$ is derived from  \eqref{decomposition2}.
 \end{remark}

  \begin{remark} We observe that infinite graphs in our framework are essentially ``acyclic'', meaning the probability that any given node will be contained in a cycle of edges of any fixed length will be zero. This property is unlike what we observe in real financial networks. Since infinite graphs are almost everywhere tree-like, one can understand that the independence conditions \eqref{decomposition2} are part of a hierarchy of similar conditions for edges and nodes in general position.  \end{remark}

\subsection{Balance Sheets}
To build a financial network with full accounting information, consistent with a given skeleton graph, one specifies the external assets $Y_v$ and external liabilities $Z_v$ for each node $v$, and for each edge $\ell$ of the network, an exposure size or weight $w_\ell$. All these quantities are positive. From this one defines the {\it net worth} or {\it buffer} of a node $v$ to be
\begin{equation}
\label{networth}
\gamma_v=Y_v+\sum_{\ell\in\Cl^-_v} w_\ell-Z_v-\sum_{\ell\in\Cl^+_v} w_\ell\; .
\end{equation}
We will always assume that the system is initially in a ``cascade equilibrium'' (or ``equilibrium'' for short) in which all banks are solvent, which means that $\gamma_v>0$ at every node $v$. Thus $\gamma_v$ can be thought of as a buffer that keeps the bank solvent when subjected to balance sheet shocks up to a certain size.

The cascade dynamics in the GK framework do not depend on full accounting information, but only on the partial information
\begin{equation}
\label{balancesheet}
\{\gamma_v,v\in\Cv\}\cup\{w_\ell, \ell\in\Cl\}.
\end{equation}
We adopt a deterministic rule for which buffers may depend on the node type $(j,k)$ but the edge weights depend only on the in-degree $\indeg(\ell)$:
\begin{equation}
\label{decorations}\left\{
\begin{array}{ll }
   \gamma_v=\gamma_{jk},   &  v\in\Cv_{jk}  \\
    w_\ell=w_{j},   &  \ell\in\Cl_{kj}
\end{array}
\right.\; .
\end{equation}  For the original GK model described in \cite{GaiKapa10}, which we will call the GK specification, the following choices are made:
\[ \gamma_{jk}=\gamma:=0.035;\quad w_j=\frac1{5j} \; ,\]
but the analytical results of that paper clearly hold for general prescriptions of the form \eqref{decorations}.

\begin{figure}[h]
\centering
\includegraphics[scale=.7]
{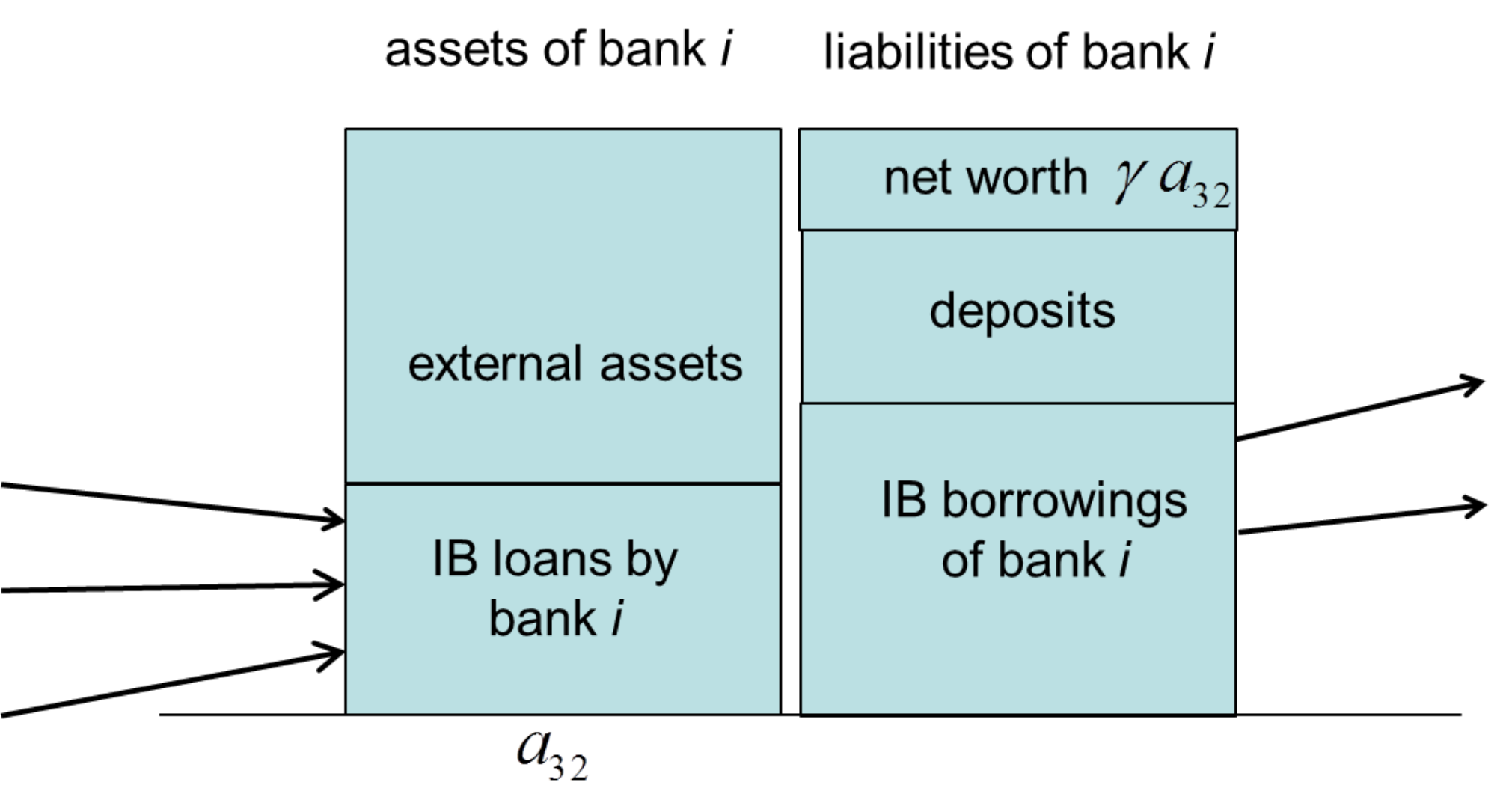}
\caption{Schematic balance sheet of banks in the $(j,k)=(3,2)$ class.}\label{figS2}
\end{figure}

\subsection{Shocks and the Solvency Condition}
Insolvencies arise in a system initially in equilibrium only when a shock hits that is hard enough to cause at least one node to suffer a loss larger than its buffer $\gamma_v$. For simplicity, we suppose that such an initial shock to our system causes an initial set $\Cw_0\subset\Cv$ of nodes to become insolvent (for example by hitting their external assets), but leaves other banks' balance sheets unchanged.   The set $\Cw_0$ is drawn randomly, with the fraction of type $(j,k)$ nodes that are defaulted denoted by
\[  \rho^0_{jk}:=\mathbb{P}[v\in\Cw_0 | v\in\Cv_{jk}]\; .\]

Under the GK ``zero recovery'' assumption that an insolvent bank can pay none of its interbank credit obligations, each insolvent node $v\in\Cw_0$ triggers all its out-edges to have zero value. This triggering of edges to default is an instance of what we call an ``edge update'' step of the cascade: corresponding to any default node set $\Cw$ there is a default edge set $\Cm\subset\Cl$ defined by the condition $\ell\in\Cm$ if and only if $v^+_\ell\in\Cw$.

Each such defaulted edge $\ell$ now transmits a maximal shock $w_\ell$ to the asset side of the balance sheet of its in-node $v^-_\ell$ (the creditor bank). If all balance sheets are determined by the reduced accounting information  $\{\gamma_{jk}\}\cup\{w_j\}$, then when $\Cm$ is a set of defaulted edges, the solvency condition on a node $v\in\Cv_{jk}$ is\footnote{The indicator function ${\bf 1}_A$ of any set $A$ is the random variable that is $1$ on the set and $0$ on its complement.}
\[ \gamma_{jk}> \sum_{\ell\in\Cl^-_v}{\bf 1}_{\{\ell\in\Cm\}} \,w_j \;  .
\]
This triggering of nodes to default we call a ``node update'' step of the cascade: corresponding to the default edge set $\Cm$ there is a default node set $\Cw'$ defined by the condition $v\in\Cw'$ if and only if
\be\label{nodeupdate}
\#\{\Cl^-_v\cap\Cm\}\le  M_{jk}:=\lceil \gamma_{jk}/w_j\rceil  \; ,
\ee
where $(j,k)$ is the type of $v$. Here $\lceil x\rceil $ denotes the ``ceiling'' function, i.e. the smallest integer greater than or equal to  $x$, and so $M_{jk}$ is the threshold for the number of defaulted in-edges that a type $(j,k)$ node can sustain without itself defaulting.

To summarize, for $N<\infty$ and $N=\infty$, a specification $(N,P,Q,\gamma,w,\rho^0)$ of an extended GK financial system is the following information: (i) the ``skeleton'', a ${\cal G}(N,P,Q)$  random directed graph defined by the probabilities $P_{jk}, Q_{kj}$ over all node and edge types; (ii) a reduced accounting set, denoted by $\{\gamma_{jk}\}\cup\{w_j\}$ and (iii) an initial shocked set $\Cw_0$ with the default probabilities $\rho^0_{jk}$ for each node type and the corresponding defaulted edge set $\Cm_0$.

Given any realization of an extended GK financial system  so specified, the complete default cascade will be a deterministic alternating sequence of node and edge updates (finite if $N$ is finite), beginning with the initial shocked set $\Cw_0$ and its corresponding edge set $\Cm_0$.
When $N=\infty$, the cascade can be fully resolved, and the expected fraction of total defaulted nodes and edges (and other statistics) can be determined by the inductive analysis given in the next section.

\section{Default Cascade Steps}\label{sec:cascade}
Given any realization of an extended GK financial system $(N,P,Q,\gamma,w,\rho^0)$ as specified above, with an initial shocked set $\Cw_0$  and the corresponding edge set $\Cm_0$, the default cascade can be thought of as a sequence of updates:
\[\begin{pmatrix}
   \Cw_0 \\
   \Cm_0
\end{pmatrix}\to \begin{pmatrix}
   \Cw_0\cup\Cw_1 \\
   \Cm_0    \cup\Cm_1
\end{pmatrix}\to \begin{pmatrix}
   \Cw_0\cup\Cw_2 \\
   \Cm_0    \cup\Cm_2
\end{pmatrix}\dots \to \begin{pmatrix}
   \Cw_0\cup\Cw_n \\
   \Cm_0    \cup\Cm_n
\end{pmatrix}\to\dots\]
Note that the above set unions are assumed to be disjoint, since we prefer to distinguish the initial default sets $ \Cw_0,
   \Cm_0 $ from the sets of ``newly defaulted'' nodes and edges. Inductively, we have increasing sequences of sets:
     \begin{eqnarray}
\Cw_n&:=&\mbox{defaulted nodes not in $\Cw_0$ ``triggered'' by edges in $\Cm_0    \cup\Cm_{n-1}$}\\
\Cm_n&:=&\mbox{defaulted edges not in $\Cm_0$ ``triggered'' by nodes in $\Cw_0\cup\Cw_{n}$} .
\end{eqnarray}
When $N=\infty$ these random default sets $\Cw_n,\Cm_n$ define probabilities for $n=0,1,2,\dots$
\begin{eqnarray}
\rho^n_{jk} & := &\mathbb{P}[v\in\Cw_n | v\in\Cv_{jk}] \label{rhondef} \\
\sigma^n_{kj} & := & \mathbb{P}[\ell\in\Cm_n | \ell\in\Cl_{kj}] \label{signdef}\; .
\end{eqnarray}

 Now the set $\Cm_0$ is determined from $\Cw_0$ by an edge update step. Similarly, for each $n\ge 1$ the set  $\Cm_0    \cup\Cm_n$ is determined from $ \Cw_0\cup\Cw_n$ by an edge update step. In all these cases the probabilities $\sigma^n_{kj}$  are determined by the following general lemma.
\begin{lemma}
\label{edgelemma} (Edge update) Let $N=\infty$ and suppose $\Cw\subset\Cv$ denotes a set of defaulted nodes and for all $j,k$, $\rho_{jk}:=\mathbb{P}[v\in\Cw | v\in\Cv_{jk}]$. If the corresponding set of defaulted edges is denoted $\Cm\subset\Cl$ then for all $j,k$
\[
\sigma_{kj}:=\mathbb{P}[\ell\in\Cm|\ell\in\Cl_{kj}]=\frac{\sum_{j'}(\rho_{j'k}P_{j'k})}{P^+_{k}}\; .
\]

\end{lemma}

\noindent{\bf Proof:\ }
When $N=\infty$ we may use the decomposition \eqref{decomposition2} and find
\beqq
\mathbb{P}[\ell\in\Cm|\ell=\Cl_{kj}]&=&\sum_{j'}\mathbb{P}[v^+_\ell\in\Cw\cap\Cv_{j'k}|\ell=\Cl_{kj}]\\
&=&\sum_{j'}\mathbb{P}[v\in\Cw|v\in\Cv_{j'k}]\ \mathbb{P}[v^+_\ell\in\Cv_{j'k}|\ell=\Cl_{kj}]\\&=&\sum_{j'}\rho_{j'k}\mathbb{P}[j_v=j'|k_v=k]\\
&=&\frac{\sum_{j'}(\rho_{j'k}P_{j'k})}{P^+_{k}}\ .
\eeqq
\qed

\medskip

Given any set $\Cm$ of defaulted edges, each of the nodes (excluding the initially defaulted nodes in $\Cw_0$) recomputes its balance sheet and the node update step leads to a subset $\Cw\subset\Cv\setminus\Cw_0$ of defaulted nodes determined by  $\Cm$. We separate out the originally defaulted nodes $\Cw_0$ since these were not triggered by defaulted edges. The probabilities associated to $\Cw$  are characterized by the following result.

\begin{lemma}
\label{nodelemma} (Node update) Let $N=\infty$ and suppose $\Cm\subset\Cl$ denotes a set of defaulted edges with associated probabilities $\sigma_{kj}:=\mathbb{P}[\ell\in\Cm | \ell\in\Cl_{kj}]$.  Then for all $j$, $a_j$ defined to be $\mathbb{P}[\ell\in\Cm|j_\ell=j]$ is given by
\[ a_j=\frac{\sum_k(Q_{kj}\sigma_{kj})}{Q^-_j} \; .\]
If the corresponding subset of $\Cv\setminus\Cw_0$ of defaulted nodes triggered by $\Cm$ is denoted $\Cw$ then for all $j,k$, $\rho_{jk}:=\mathbb{P}[v\in\Cw | v\in\Cv_{jk}]$ is given by
\[
\rho_{jk}=(1-\rho^0_{jk})\sum_{m= M_{jk}}^j\dbinom{j}{m} a_j^m(1-a_j)^{j-m}\]
where the default thresholds are defined  as in \eqref{nodeupdate}  by $M_{jk}=\lceil \gamma_{jk}/w_j\rceil$.
\end{lemma}

\noindent{\bf Proof:\ } First we compute that
\be\label{edgeprob}
a_j=\mathbb{P}[\ell\in\Cm|j_\ell=j]=\frac{\sum_{k}\mathbb{P}[\ell\in\Cm\cap\Cl_{kj}]}{\mathbb{P}[j_\ell=j]}=\frac{\sum_{k}(Q_{kj}\sigma_{kj})}{Q^-_j}  \; .
\ee
Note that $\mathbb{P}[\ell\in\Cm|v^-_\ell\in\Cv_{jk}]=\mathbb{P}[\ell\in\Cm|j_\ell=j]$, that
\be\mathbb{P}[v\in\Cw | v\in\Cv_{jk}]=\mathbb{P}[v\in\Cw | v\in\Cv_{jk}\setminus\Cw_0]\ \mathbb{P}[v\notin\Cw_0]=(1-\rho^0_{jk})\mathbb{P}[v\in\Cw | v\in\Cv_{jk}\setminus\Cw_0]\; ,
\label{pcond}\ee
and that a node $v\in\Cv_{jk}\setminus\Cw_0$ will be in default if and only if at least $M_{jk}$ in-edges to $v$ are in $\Cm$.  By the independence structure of the network for $N=\infty$, the random variables ${\bf 1}_{\ell\in\Cm}$ for all $\ell\in\Cl^-_v$, under the condition that $v\in\Cv_{jk}\setminus\Cw_0$,  are a collection of $j$ identical independent Bernoulli random variables with probability $a_j$. Putting these facts together gives \[
\mathbb{P}[v\in\Cw | v\in\Cv_{jk}\setminus\Cw_0]=\sum_{m= M_{jk}}^j\dbinom{j}{m} a_j^m(1-a_j)^{j-m}
\]
which combined with \eqref{pcond}  leads to the required result.

 \qed

\medskip

 Using these lemmas and the definitions \eqref{rhondef},\eqref{signdef}, it is straightforward to piece together the steps of the default cascade and obtain the main result of the paper.
\begin{proposition} Consider the infinite extended GK financial network $(N=\infty,P,Q,\gamma,w,\rho^0)$. For $n=0, 1,2,\dots$, let $\vec{a}^{(n)}=\{a^{(n)}_j\}$ denote the probabilities
$\mathbb{P}[\ell\in\Cm_0\cup\Cm_n|j_\ell=j]$.  Then
\begin{enumerate}
\item For $n=0$ we have
\beq\label{sig0}
\sigma^0_{kj}&=&\frac{\sum_{j'}(\rho^0_{j'k}P_{j'k})}{P^+_{k}}\\
 a^{(0)}_j&=&\frac{\sum_k(Q_{kj}\sigma^0_{kj})}{Q^-_j}.
\label{a0} \eeq
  \item For $n=1,2,\dots$, the quantities $\rho^n,\sigma^n,a^{(n)}$ satisfy the recursive formulas
  \beq
  \label{rhon}\rho^n_{jk}&=&(1-\rho^0_{jk})\sum_{m= M_{jk}}^j\dbinom{j}{m} (a^{(n-1)}_j)^m(1-a^{(n-1)}_j)^{j-m}\\
\label{sign} \sigma^n_{kj}&=&\frac{\sum_{j'}(\rho^n_{j'k}P_{j'k})}{P^+_{k}}\\
 \label{an} a^{(n)}_j&=&\frac{\sum_k(Q_{kj}(\sigma^0_{kj}+\sigma^n_{kj}))}{Q^-_j}\; ,
 \eeq
 where $M_{jk}=\lceil \gamma_{jk}/w_j\rceil$.
  The total probability for defaulted $(j,k)$ nodes at step $n$ is $\rho^0_{jk}+\rho^n_{jk}$ and the total probability for defaulted $(k,j)$ edges at step $n$ is  $\sigma^{0}_{kj}+\sigma^{n}_{kj}$.
\item The new probabilities $\vec{a}^{(n)}=\{a^{(n)}_j\}$ are a function $G(\vec{a}^{(n-1)})$ which is explicit in terms of the specification $(N,P,Q,c,w,\rho^0)$. \end{enumerate}
\end{proposition}

\noindent{\bf Proof:\ } In Part 1, \eqref{sig0} follows from Lemma \ref{edgelemma} and \eqref{a0} from Lemma \ref{nodelemma}. Part 2 is immediate from the same  two lemmas, while Part 3 is simply a composition of \eqref{rhon}, \eqref{sign}, \eqref{an}.

 \qed

\medskip

\section{The Cascade Condition}\label{sec:casccond}
The size of global cascades in an extended GK financial network with $N=\infty$ has essentially been reduced to solving the fixed point equation
\be\label{fix} \vec{a}=G(\vec{a})
\ee
by iteration of the mapping $G$. Scalar equations of this sort, giving the expected size of cascades on directed  networks, have been previously derived in various contexts \cite{Gleeson08b,AminContMinc10}. In \cite{Gleeson08b}, the main focus is on percolation-type phenomena (see also the undirected networks case \cite{Gleeson08a}), while \cite{AminContMinc10} considers more complicated dynamics but takes the limit $\rho^0\to0$. The case considered in \cite{GleHurMelHac11}, where initial default fractions can be different for each $(j,k)$ class, has not, to our knowledge, been considered previously. In the current work, we include for the first time (through $Q_{kj}$) the effect of non-trivial correlations between the degrees of nodes at either end of a randomly chosen edge.

As a consequence of the Knaster-Tarski Theorem,  equation \eqref{fix} always has at least one solution $\vec{a}^{\infty}$ and this will be a vector of probabilities  $a^{\infty}_j\in[0,1]$ for all $j$. To see this one observes that $G$ is a monotone mapping from the complete lattice $[0,1]^{\mathbb{Z}_+}$ onto itself, under the partial ordering relation defined by $\vec{a}\le\vec{b}$ if and only if $a_j\le b_j$ for all $j\in\mathbb{Z}_+$, that is, $G(\vec{a})\le G(\vec{b})$ whenever  $\vec{a}\le\vec{b}$. This is enough to ensure the existence of at least one fixed point on the set $[0,1]^{\mathbb{Z}_+}$.

One important question is to consider initial points $\epsilon \vec{a}$ for small $\epsilon>0$ and ask whether the fixed points $\vec{a}_\infty(\epsilon)$ obtained this way are of order $\epsilon$ or of order $1$ as $\epsilon\to 0$. In other words, what is the ``cascade condition'' that determines if an infinitesimally small seed fraction will grow to a large-scale cascade? This depends on the spectral radius of the derivative matrix $D=\{D_{jj'}\}$ with $D_{jj'}=\partial G_j/\partial a_{j'}|_{\vec{a}={\bf 0},\rho^0=0}$. Recalling that the spectral radius of a matrix $D$, $\|D\|:=\max_{\vec{a}:\|\vec{a}\|=1}\|D\vec{a}\|$, is the magnitude of the largest eigenvalue of $D$,  one can see by a version of the Perron-Frobenius Theorem for non-negative infinite matrices, that the fixed points $\vec{a}_\infty(\epsilon)$ will be $O(\epsilon)$   if $\|D\|<1$ and  will be $O(1)$  if   $\|D\|>1$. Such a  ``cascade condition'' plays a role in systemic risk analogous to the basic reproduction number  $R_0$ in epidemiology.  In our framework, the derivatives $D_{jj'}$  turn out to be easy to calculate and we find:
\begin{proposition} The infinite extended GK financial network $(N=\infty,P,Q,\gamma,w,\rho^0)$  satisfies the cascade condition, that is, any infinitesimal seed will trigger a large scale cascade almost surely, if the spectral radius $\|D\|>1$ where
\be\label{cascadecond}
D_{jj'}=\sum_k\frac{j'Q_{kj}P_{j'k}{\bf 1}_{\{\gamma_{j'k}\le w_{j'}\}}}{Q_j^-P_k^+}\; .
\ee
If $\|D\|<1$, then almost surely the network will not exhibit large scale cascades for any infinitesimal seed.
\end{proposition}

In Section~\ref{sec:num},
 we shall see that the cascade condition is indeed a strong measure of systemic risk in finite simulated networks. One can check that in the setting of independent edge probabilities, the mapping $G$ reduces to a scalar function, a result that has been derived in a rather different fashion in \cite{GaiKapa10}. They extend Watts' \cite{Watts02} percolation theory approach from his work on undirected networks to the case of directed nonassortative networks. We will see in the next section that the percolation approach to the cascade condition extends further to our directed assortative networks.

We can understand the cascade condition more clearly by introducing the notion of {\it vulnerable node}, that is any node that becomes insolvent if any one of its debtors defaults. In our specifications, a $(j,k)$ node is thus vulnerable if and only if $\gamma_{jk}\le w_{j}$. The matrix element $D_{jj'}$ has a simple explanation that gives more intuition about the nature of the cascade condition: it is the expected number of edges $\ell'$ with $j_{\ell'}=j'$ that connect through a vulnerable node to an edge $\ell$ with $j_{\ell}=j$.
Then for small values of $\vec{a}$, one has a linear approximation for the change in  $\vec{a}$ in a single cascade step:
\be a^{m+1}_j-a^{m}_j= \sum_{j'} D_{jj'} \ (a^m_{j'}-a^{m-1}_{j'}) + O(\|a\|^2)\ .
\ee
The  condition for a global cascade starting from an infinitesimal seed is that the matrix $D$ must have an expanding direction, i.e. an eigenvalue bigger than 1.

\section{Frequency of global cascades and the giant vulnerable cluster}\label{Frequency}
The previous argument does not tell us directly about the frequency of global cascades. However, it is well-known \cite[Chapter 13.11]{Newman10} that the frequency of global cascades in infinite random graphs is given by the fractional size of the so-called in-component associated to the giant vulnerable cluster.

To facilitate the discussion we make the following further definitions \begin{itemize}
\item $\Cv_{(v)}$ is the set of vulnerable nodes.
  \item $\Cs_s$ is the giant strongly connected set of vulnerable nodes (the ``giant vulnerable cluster'');   \item $\Cs_i$ is the set of (possibly not vulnerable) nodes that are forward connected to $\Cs_s$ by a path of vulnerable nodes (the ``in-component'' of the giant vulnerable cluster);
  \item $\Gamma_{jk}={\bf 1}_{\{\gamma_{jk}\le w_j\}}$.
\end{itemize}

	We are interested in the following probabilities $\vec b=(b_k),\ b_k:=\mathbb{P}[v\notin \Cs_i  |k_v=k]$ and note that $v\in \Cs_i^c$ (i.e. the complement of $\Cs_i$) is equivalent to  the condition that all the downstream nodes $v^-_\ell, \ell\in\Cl^+_v$ are in the set $(\Cs_i^c\cap\Cv_{(v)})\cup\Cv_{(v)}^c$. Thus, letting $v'$ denote any node one edge downstream from $v$, one has:
\be
b_k =(c_k)^k\ ,
\ee
where
\[ c_k= \sum_{j',k'} \mathbb{P}[v'\in (\Cs_i^c\cap\Cv_{(v)})\cup\Cv_{(v)}^c |v'\in\Cv_{j'k'},k_\ell=k]\mathbb{P}[v'\in\Cv_{j'k'}|k_\ell=k]\ .
\]
Next note
\beqq
&&\mathbb{P}[v'\in (\Cs_i^c\cap\Cv_{(v)})\cup\Cv_{(v)}^c) |v'\in\Cv_{j'k'},k_\ell=k]=\Gamma_{j'k'}b_{k'}+(1-\Gamma_{j'k'})\\
&&\mathbb{P}[v'\in\Cv_{j'k'}|k_\ell=k]=\frac{P_{j'k'}Q_{kj'}}{P^-_{j'}Q^+_k}
\eeqq
and thus
\be c_k=\sum_{j',k'} \left(\Gamma_{j'k'}b_{k'}+(1-\Gamma_{j'k'})\right)\frac{P_{j'k'}Q_{kj'}}{P^-_{j'}Q^+_k}\ .
\ee
Since $b_k=(c_k)^k$ it follows that  $\vec c=(c_k)$ satisfies the fixed point equation $\vec c=h(\vec c)$ where for any sequence $c=(c_k)$
\be h_k(c)= \sum_{j',k'} \left(\Gamma_{j'k'}(c_{k'})^{k'}+(1-\Gamma_{j'k'})\right)\frac{P_{j'k'}Q_{kj'}}{P^-_{j'}Q^+_k}\ .
\ee

Note that the equation $\vec c=h(\vec c)$  has a trivial fixed point $\vec e=(1,1,\dots)$ that corresponds to the set $\Cs_i$ having probability zero. We  now verify that the cascade condition $\|D\|>1$ is equivalent to the condition that $e$ is an unstable fixed point, in which case there will be a nontrivial fixed point $0\le \vec c_\infty < \vec e$. A sufficient (and almost necessary) condition for $e$ to be an unstable fixed point is that $\|\tilde D\|>1$ where the derivative $\tilde D_{kk'}=(\partial h_k/\partial c_{k'})|_{\vec c=\vec e}$ is given by
\be \tilde D_{kk'}=\sum_{j'}\frac{k'Q_{kj'}P_{j'k'}\Gamma_{j'k'}}{Q^+_k P^-_{j'}}\ . \ee
One can verify directly that
\[ \tilde D=\left(\Lambda BA\Lambda^{-1}\right)^T, \quad D=AB
\]
for matrices
\[ A_{jk}=\frac{Q_{kj}}{Q^-_j},\ B_{kj'}=\frac{j' P_{j'k}\Gamma_{j'k}}{P^+_k},\ \Lambda_{kk'}=\delta_{kk'}kP^+_k
\]
and from this it follows that the spectral radii and spectral norms of $\tilde D$ and $D$ are equal. Hence $\|D\|>1$ if and only if $\|\tilde D\|>1$.

As long as the cascade condition is satisfied, the cascade frequency $f$ equals the fractional size of the in-component $\Cs_i$ and is given by
\be\label{frequency} f=
\sum_k \mathbb{P}[v\in \Cs_i  |k_v=k]\ \mathbb{P}[k_v=k]=\sum_k(1-c_{k,\infty})P^+_k\ .
\ee

Repeating this type of argument to determine the size of the out-component of the giant vulnerable cluster, one also obtains an upper bound on the size of the global cascade.

\section{Numerical Results}\label{sec:num}

In this section we present results from large-scale Monte Carlo simulations on random networks, and show that the analytical theory of Section~\ref{sec:cascade} for $N=\infty$ matches well to the numerical results when $N$, the number of nodes in the network, is sufficiently large.

\subsection{A Simple Random Network Model}\label{sec:Ex1}
We consider networks constructed with nodes of types $(3,3), (3,12), (12,3), (12,12)$ and edges of the same types. We fix the marginal probabilities $P^+_3=P^+_{12}=1/2$ which lead to an average degree $z=15/2$ and the marginals $Q^+_3=1/5,Q^+_{12}=4/5$. For parameters $a\in[0, 1/2]$ and  $b\in[0,1/5]$ the following $P$ and $Q$ probabilities are consistent:
\begin{equation}
\label{PQ}
\begin{pmatrix}
  P_{3,3}    &  P_{3,12}  \\
  P_{12,3}    &  P_{12,12}
\end{pmatrix}=\begin{pmatrix}
 1/2-a     &a    \\
   a   &  1/2-a
\end{pmatrix};\quad \begin{pmatrix}
  Q_{3,3}    &  Q_{3,12}  \\
  Q_{12,3}    &  Q_{12,12}
\end{pmatrix}=\begin{pmatrix}
 1/5-b     &b    \\
   b   &  4/5-b
\end{pmatrix}\; .
\end{equation}

We first fix the value of $a$ to be $0.5$, which means that the in- and out-degrees of all nodes are negatively correlated: nodes with in-degree 3 have out-degree 12, and vice versa. We examine three different values of the parameter $b$: the independent connections case $b=0.16$, the (almost) maximally assortative case $b=0.01$ and the (almost)  maximally disassortative case $b=0.19$.  Note that the independent edge condition has been assumed in all previous work on such problems. We also note that with $b=0$, edges are maximally assortative and link nodes of out-degree 3 to nodes of in-degree 3 only, and nodes of out-degree 12 to nodes of in-degree 12 only. In this case, the network falls into two disjoint pieces.

The balance sheet quantities are those of \cite{GaiKapa10} (except for the percentage net worth $\gamma$, which we vary over the range $0\%$ to $10\%$), while the initial shock distribution is taken to be
 $\rho^0_{jk}=1/N$ for all types $(j,k)$, corresponding to the shocking of a single, randomly-chosen, bank.

Generating a finite size skeleton network with prescribed probabilities following the steps outlined in section \ref{sec:skel} always encounters the difficulty of ensuring that the consistency condition for wiring the network is exact. The naive approach would be to accept any randomly drawn set of node types $\{(j_n,k_n)\}_{n=1,\dots,N}$ only if the consistency condition is satisfied: one can easily see that even for our simple probability specification the rejection rate will be extremely high, and the method very slow. Instead one in practice adopts some approximation that we call ``clipping'': given an inconsistent labelling, one selects random pairs of nodes, and ``rewires'' a small number of times to ensure the consistency conditions. Such a procedure may create a small bias that vanishes as $N\to\infty$. 

\begin{figure}[h]
\centering \vspace{-1.3in}
\includegraphics[scale=0.9]
{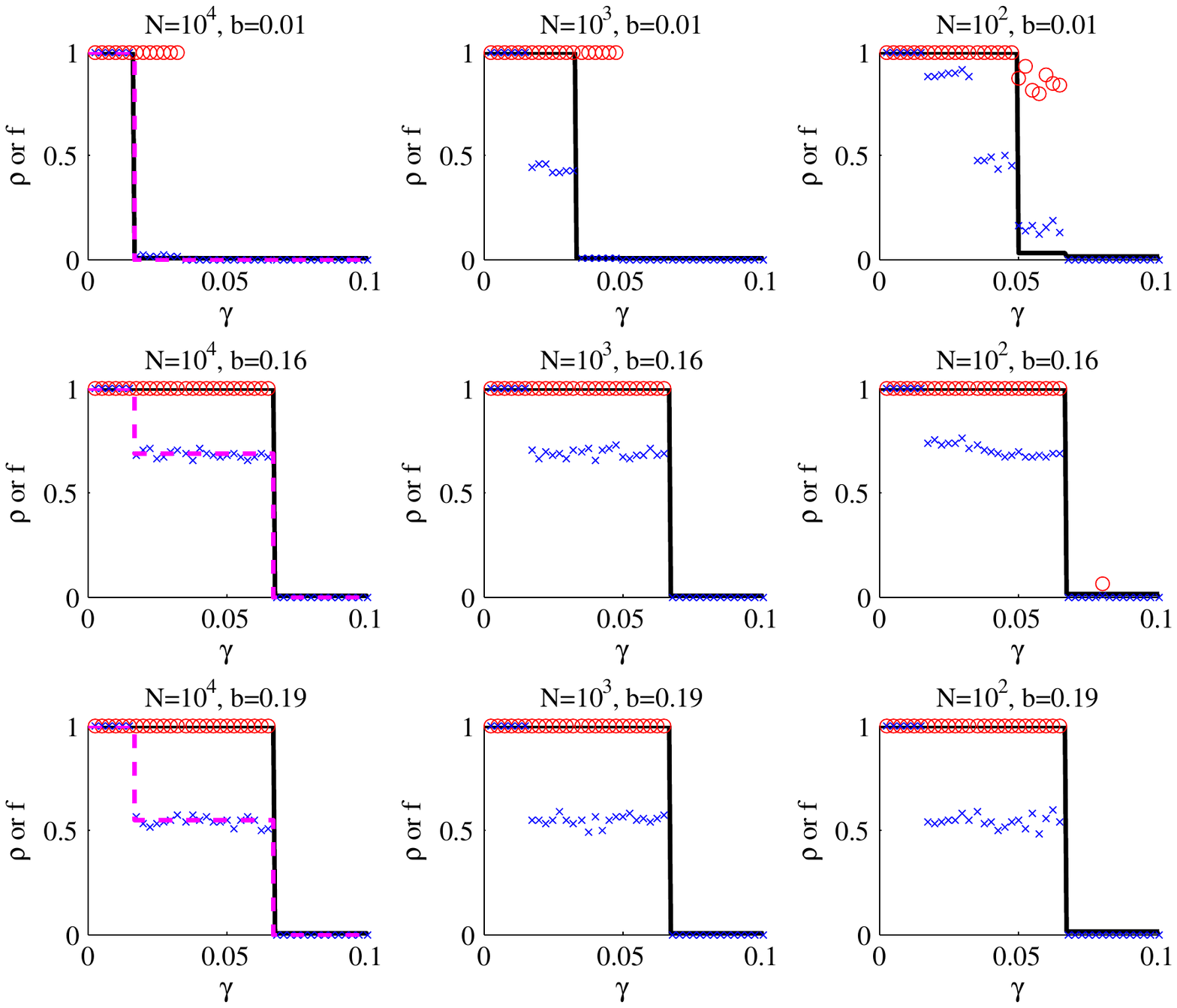}
\vspace{-2.5in}\caption{Numerical simulation results (symbols) and theoretical results (curves) for the random network model of Equation \eqref{PQ}, on networks of $N$ nodes with parameter $a=0.5$, as functions of the net worth $\gamma$. The average size and frequency of global cascades in simulations are shown by red circles and blue crosses, respectively. Theoretical results for the expected cascade size (black solid curve) are from Section 3; those for the frequency of cascades (dashed magenta curve) are from Section 5. Each column shows results for a different network size $N$, and the parameter $b$ takes a different value on each row of the figure.} \label{figR1}

\end{figure}

Figure~\ref{figR1}  compares theory curves for cascade size (found by iterating equations (\ref{sig0})--(\ref{an}) to convergence) as well as cascade frequency (given by \eqref{frequency}) with results from numerical simulations on random networks with $N=10^4, 10^3$ and $10^2$ nodes. The nodal correlation parameter is fixed at $a=0.5$, while the edge correlation parameter takes the values $b=0.01, 0.16, 0.19$. Results are plotted as functions of the percentage net worth parameter $\gamma$. In each case, 500 realizations are used to find the extent of global cascades (a global cascade is defined, similarly to \cite{GaiKapa10,GleHurMelHac11}, as one in which more than 
 5\% of nodes default), and the frequency with which such global cascades occur. As expected, the analytical approach accurately predicts the size of the global cascades. Some discrepancies may be noted in Figure~\ref{figR1}, where the theory does not predict some global cascades, but note that these occur with only very small frequencies. 

The cascade condition (\ref{cascadecond}) predicts that the critical values of the buffer parameter $\gamma$ are: $\gamma_c=0.067$ for the parameters of Figure~\ref{figR1}(a), and $\gamma_c=0.017$ for the case of Figure~\ref{figR1}(b). These values match very accurately to the locations of the dramatic transitions in the theory curve (and in the expected size of cascades in numerical experiments): for buffer values in excess of $\gamma_c$ global cascades are extremely rare, while for $\gamma$ values  less than $\gamma_c$ the entire financial system is likely to fail following a single bank's default.  These result indicate the potential usefulness of the cascade condition as a measure of systemic risk.


\begin{figure}[h]
\centering\vspace{-1.5in}\includegraphics
{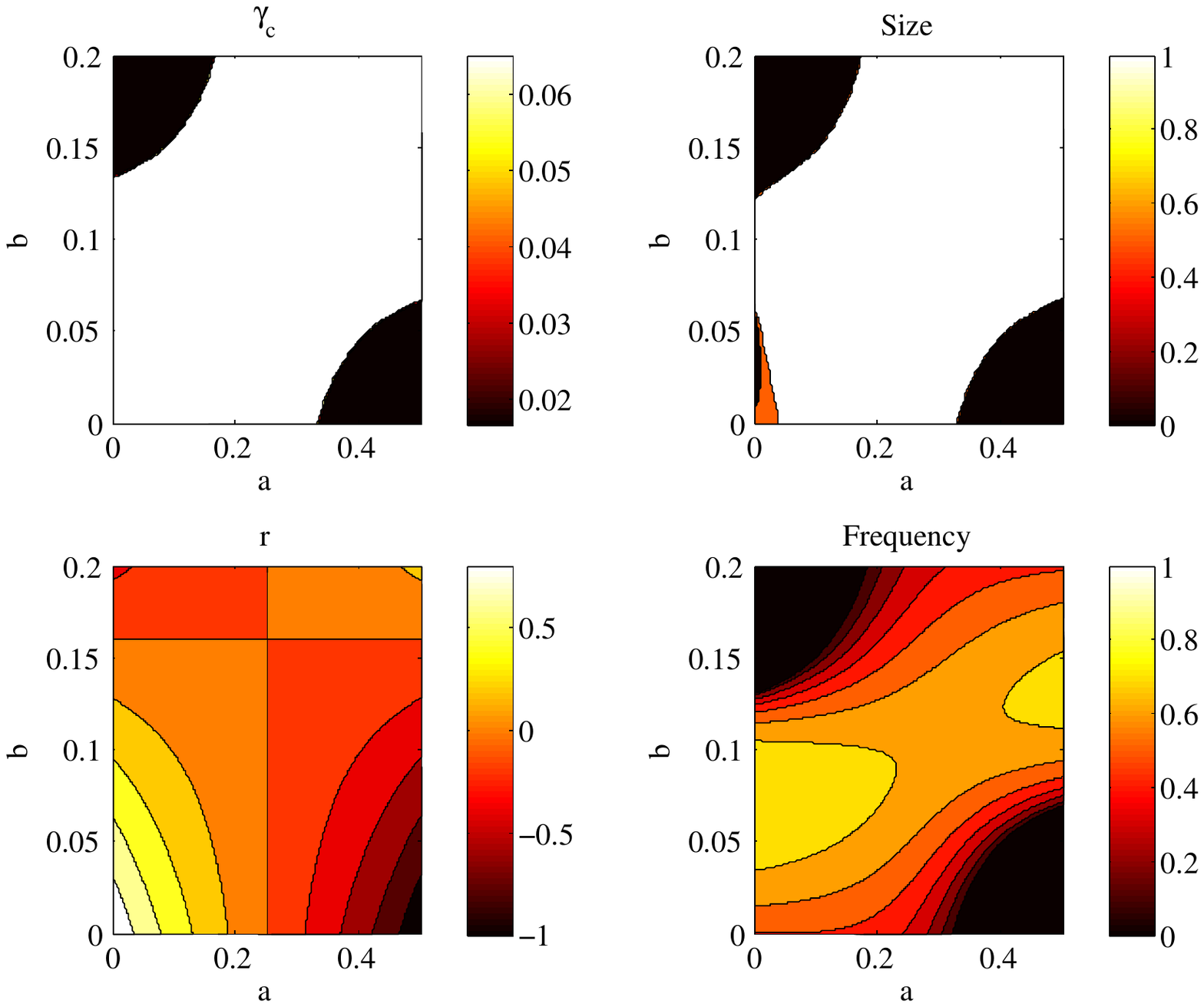}
\vspace{-3in}\caption{Theoretical joint dependences on the graph parameters $(a,b)$. Top left: critical $\gamma$ value for the random network model of Equation \eqref{PQ}.
Top right: Expected size of cascades (from Section 3) when $\gamma=0.05$ and $\rho^0_{j k}=10^{-4}$.  Bottom left: the graph assortativity parameter $r$.
Bottom right: Frequency of cascades (from Section 5) when $\gamma=0.05$.}
\label{fig_draw_SIFIN_fig2}
\end{figure}

We consider  in Figure~\ref{fig_draw_SIFIN_fig2} the joint dependences on $a,b$ of various theoretical quantities in the large $N$ limit. In the top figures, the critical value of $\gamma$ and cascade size are seen to be discontinuous, and not directly related to edge-assortativity (parametrized by $b$). On the other hand
(see bottom figures), the frequency of cascades is continuously varying, and does appear to correlate somewhat with the graph assortativity coefficient $r$ given by \eqref{graphassortcoef}.

\subsection{Another Simple Random Network Model}

Now we have shown that the infinite $N$ theory meets our expectations, we can further explore the implications of the analytical method. Since the specification of extended GK networks has many components, one must be rather careful in the questions one wishes to address: we choose here to try by means of a simple network specification to shed some additional light on the role the assortative properties of a network play in its susceptibility to systemic risk as measured by the cascade frequency.

We consider stylized networks with many small banks and a few large banks. The set of node types will be $\{ (2,2), \ (4,4), \ (8,8),\ (16,16)\}$ with a diagonal node probability matrix:
\[ P:=(P_{jk})=\mbox{diag}(8,\ 4,\ 2,\ 1)/15\ .\]
The following edge probability matrices $Q=(Q_{kj})$ are consistent with $P$:
\beqq Q^{(1)}&=&\frac14\begin{pmatrix}
 1     & 0&0&0   \\
    0  & 1&0&0\\
    0&0&1&0\\
    0&0&0&1
\end{pmatrix},\
 Q^{(2)}=\frac14\begin{pmatrix}
 0     & 1&0&0   \\
    1  & 0&0&0\\
    0&0&0&1\\
    0&0&1&0
\end{pmatrix},\ \\
 Q^{(3)}&=&\frac14\begin{pmatrix}
 0     & 0&1&0   \\
   0  & 0&0&1\\
    1&0&0&0\\
    0&1&0&0
\end{pmatrix},\
 Q^{(4)}=\frac14\begin{pmatrix}
 0     & 0&0&1   \\
    0  & 0&1&0\\
    0&1&0&0\\
    1&0&0&0
\end{pmatrix},\eeqq
and their convex combinations $q_1Q_1+q_2Q_2+q_3Q_3+q_4Q_4$, $q_1+q_2+q_3+q_4=1, q_i\ge 0$ span a simplex of possible edge probability matrices. We can see that as measured by $r_Q$, $Q^{(1)}$ is maximally assortative, while $Q^{(3)}$ and $Q^{(4)}$  are maximally disassortative, and the independent case is $Q^{(0)}:=[Q^{(1)}+Q^{(2)}+Q^{(3)}+Q^{(4)}]/4$.

For the remaining components of the specification we adopt the default GK balance sheet values but with $\gamma$ a variable parameter and consider  shocking a single randomly selected node (this is an infinitesimal shock in infinite volume). We then compute the critical $\gamma_c$ using the cascade formula \eqref{cascadecond},
 the cascade size from Proposition 8,
 and the default frequency using equation \eqref{frequency}.

\begin{figure}[h]
\centering\vspace{-1.7in}\includegraphics
{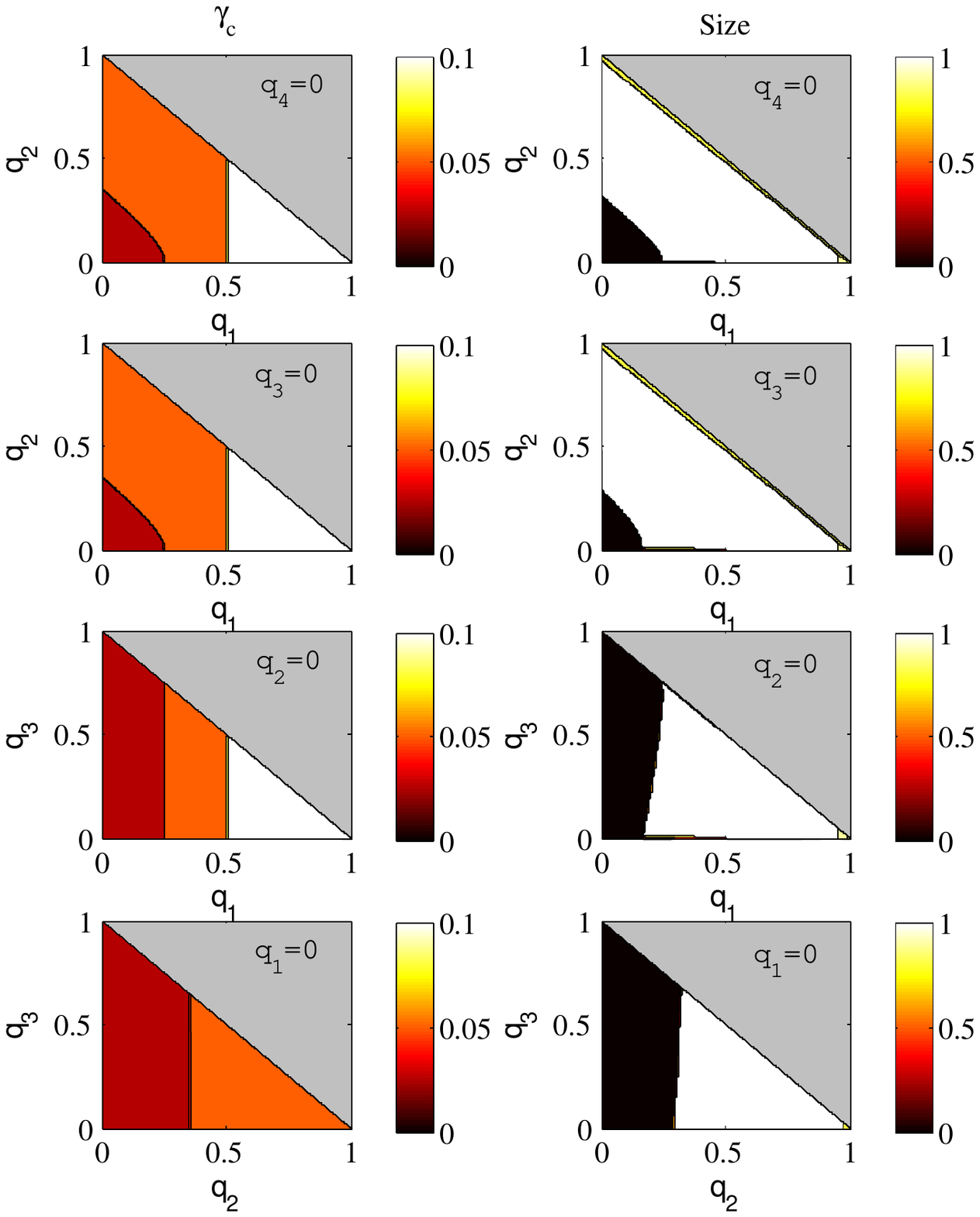}%
\vspace{-2.5in}
\caption{Critical value of $\gamma$ (left column) and expected size (right column) for cascades on the random network model of Section 6.2 with
$\gamma=0.0375$ and $\rho^0_{j k}=10^{-4}$. The triangles shown correspond to $Q$ matrices with (from top to bottom) $q_4=0$, $q_3=0$, $q_2=0$, and $q_1=0$.  }%
\label{fig_draw_SIFIN_fig3b}%
\end{figure}

\begin{figure}[h]
\centering
\vspace{-1.7in}\includegraphics
{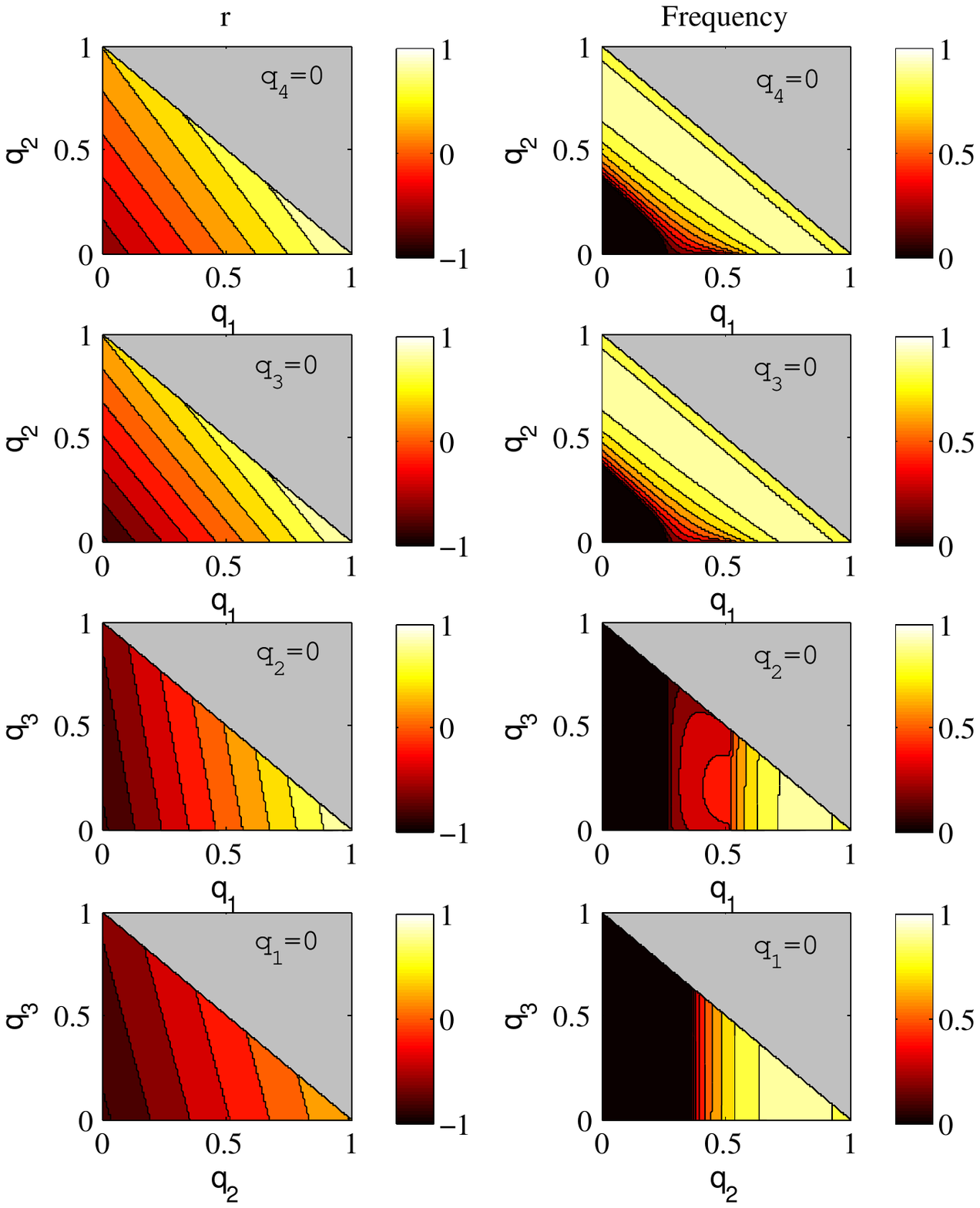}%
\vspace{-2.5in}
\caption{Graph assortativity parameter $r$ (left column) and frequency (right column)  for the same parameters as Fig. \ref{fig_draw_SIFIN_fig3b}.}%
\label{fig_draw_SIFIN_fig4}%
\end{figure}

Figure \ref{fig_draw_SIFIN_fig3b} shows how the theoretical values of $\gamma_c$ and cascade size depend on the particular  $Q$ matrix. Figure \ref{fig_draw_SIFIN_fig4} shows how the theoretical values of the graph assortativity coefficient and cascade frequency depend on $Q$. In both figures, the four rows correspond to the simplices of $Q$ matrices with $q_4=0, q_3=0, q_2=0, q_1=0$ respectively.

We see again in these networks that $r$ and $f$ vary continuously, while $\gamma_c$ and cascade size take on only discrete values. Of particular interest is the discernable covariation of $f$ with $r$. Since $r$ depends only on the skeleton graph and not on the balance sheet data, we cannot expect a one-to-one relationship between the quantities $f$ and $r$. However, we conjecture that $r$ is in some sense the best possible purely graph theoretic measure of systemic susceptibility. Heuristically, we might expect that systemic risk is lowered if, all else being equal, the network is such that the correlation between in-degrees of neighbouring nodes is lowered.

\section{Concluding Remarks}\label{sec:conclusion}
In summary, we have described here a  rigorous analytical framework which can predict the systemic risk of ``deliberately simplified models'' such as \cite{GaiKapa10}.  The qualitative type of networks one can address has been extended considerably over existing work, in particular by the inclusion of the non-independent connections between nodes. The example of Subsection~\ref{sec:Ex1} demonstrates that finite size effects do not appear to dramatically impact systemic risk as long as $N \gtrsim 100$. More subtly, we also observed that graph-assortativity, rather than edge- or node-assortativity, can strongly affect the course of contagion cascades, and hence show the importance of incorporating assortativity in numerical and analytical treatments of banking network models. Our framework  will enable extensive studies of alternative network topologies; the cascade condition and cascade frequency provide two simple and useful measures of systemic risk by which to compare different network topologies. However, the daunting range of network variables means that both analytical and numerical studies must be carefully framed to address specific issues, for example, to uncover other key determinants of systemic risk. Finally, we anticipate that future work can show how the approach described here may be further extended to include partial recovery models (such as \cite{NieYanYorAle07}) and stochastic balance sheets.


\section{Acknowledgements}
This work was funded by awards from  Science Foundation Ireland Science Foundation Ireland (06/IN.1/I366 and MACSI
06/MI/005) and from the Natural Sciences and Engineering Research Council of Canada.

\bibliographystyle{abbrvnat}

\begin{thebibliography}{12}
\providecommand{\natexlab}[1]{#1}
\providecommand{\url}[1]{\texttt{#1}}
\expandafter\ifx\csname urlstyle\endcsname\relax
  \providecommand{\doi}[1]{doi: #1}\else
  \providecommand{\doi}{doi: \begingroup \urlstyle{rm}\Url}\fi

\bibitem[Amini et~al.(2010)Amini, Cont, and Minca]{AminContMinc10}
H.~Amini, R.~Cont, and A.~Minca.
\newblock Resilience to contagion in financial networks.
\newblock Working paper, May 2010.

\bibitem[Bogu\~n\'a and Serrano(2005)]{BoguAnge05}
M.~Bogu\~n\'a and M.~A. Serrano.
\newblock Generalized percolation in random directed networks.
\newblock \emph{Phys. Rev. E}, 72:\penalty0 016106, 2005.

\bibitem[Cont et~al.(2010)Cont, Moussa, and Santos]{ContMousSant10}
R.~Cont, A.~Moussa, and E.~B. Santos.
\newblock {Network Structure and Systemic Risk in Banking Systems}.
\newblock \emph{SSRN eLibrary}, 2010.

\bibitem[Gai and Kapadia(2010)]{GaiKapa10}
P.~Gai and S.~Kapadia.
\newblock Contagion in financial networks.
\newblock \emph{Proceedings of the Royal Society A}, 466\penalty0
  (2120):\penalty0 2401--2423, 2010.

\bibitem[Gleeson et~al.(2011)Gleeson, Hurd, Melnik, and
  Hackett]{GleHurMelHac11}
J.~Gleeson, T.~R. Hurd, S.~Melnik, and A.~Hackett.
\newblock Systemic risk in banking networks without {M}onte {C}arlo simulation.
\newblock In E.~Kranakis, editor, \emph{Advances in Network Analysis and its
  Applications}, Mathematics in Industry. Springer Verlag, Berlin Heidelberg
  New York, June 2011.

\bibitem[Gleeson(2008{\natexlab{a}})]{Gleeson08a}
J.~P. Gleeson.
\newblock Mean size of avalanches on directed random networks with arbitrary
  degree distributions.
\newblock \emph{Phys. Rev. E}, 77:\penalty0 057101, 2008{\natexlab{a}}.

\bibitem[Gleeson(2008{\natexlab{b}})]{Gleeson08b}
J.~P. Gleeson.
\newblock Cascades on correlated and modular random networks.
\newblock \emph{Phys. Rev. E}, 77:\penalty0 046117, 2008{\natexlab{b}}.

\bibitem[Newman(2002)]{Newman02}
M.~E.~J. Newman.
\newblock Assortative mixing in networks.
\newblock \emph{Physic. Rev. Lett.}, 89\penalty0 (20), 2002.

\bibitem[Newman(2010)]{Newman10}
M.~E.~J. Newman.
\newblock \emph{Networks: An Introduction}.
\newblock Oxford University Press, 2010.

\bibitem[Nier et~al.(2007)Nier, Yang, Yorulmazer, and Alentorn]{NieYanYorAle07}
E.~Nier, J.~Yang, T.~Yorulmazer, and A.~Alentorn.
\newblock {Network Models and Financial Stability}.
\newblock \emph{J. Econ. Dyn. Control}, 31:\penalty0 2033--2060, 2007.

\bibitem[Upper(2011)]{Upper11}
C.~Upper.
\newblock Simulation methods to assess the danger of contagion in interbank
  markets.
\newblock \emph{J. Financial Stability}, 2011.

\bibitem[Watts(2002)]{Watts02}
D.~Watts.
\newblock A simple model of global cascades on random networks.
\newblock \emph{PNAS}, 99\penalty0 (9):\penalty0 5766--5771, 2002.

\end{thebibliography}

\end{document}